\documentclass[pre,twocolumn,showpacs,floatfix]{revtex4}
 \usepackage{graphicx}
 \usepackage{epstopdf}
 \usepackage{dcolumn}
 \usepackage{color}
 \usepackage{bm}
 \usepackage{verbatim}
 \usepackage{multirow}
 \usepackage{amsmath,amssymb,graphicx}
 \usepackage{epsfig}
 \usepackage{enumerate}
 \usepackage{array}
 \usepackage{multirow}
\newcommand{\figOne}
{
 \begin{figure*}[t]
 \begin{center}
 \includegraphics[width=2.5in,height=1.25in]{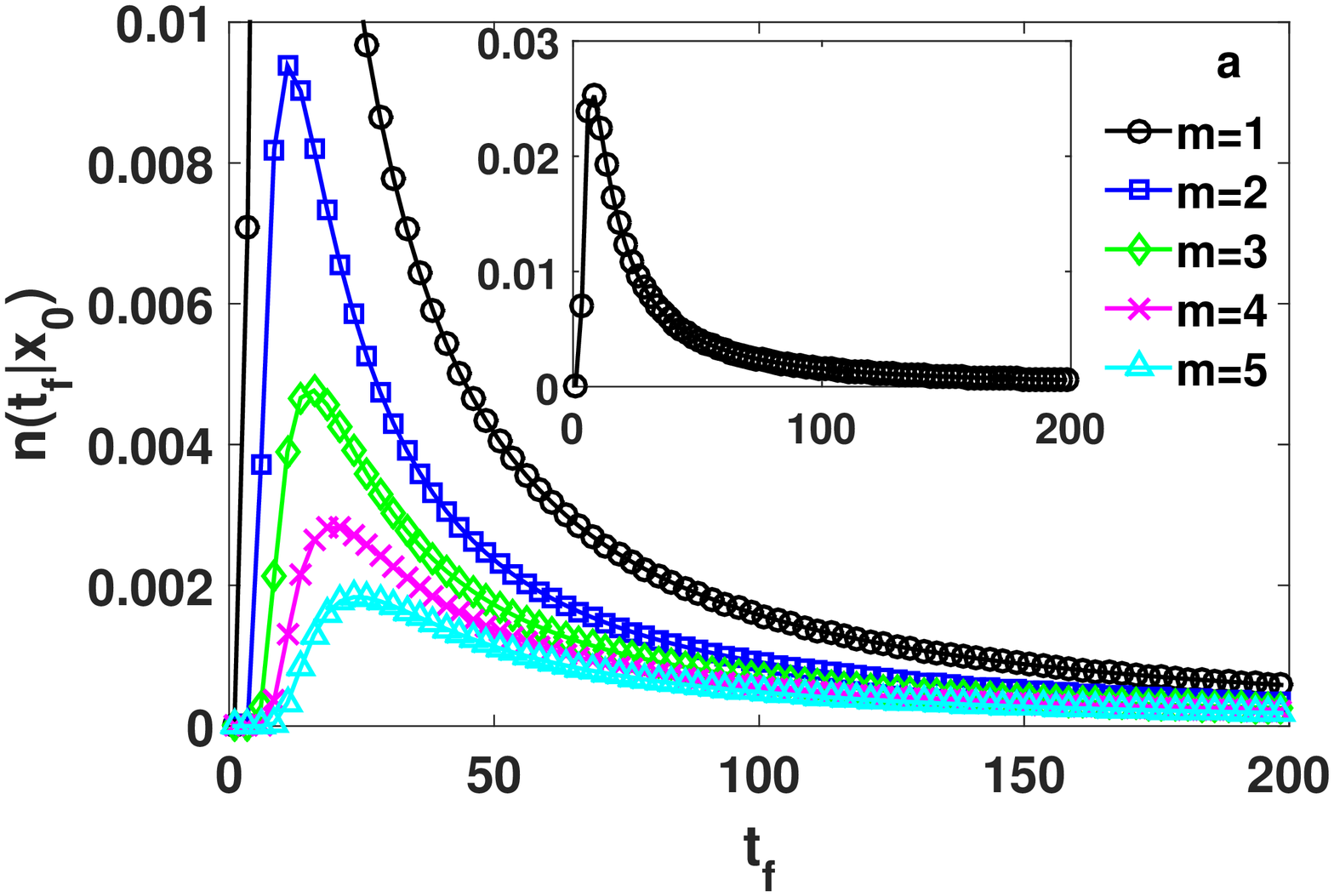}
 \includegraphics[width=2.5in,height=1.25in]{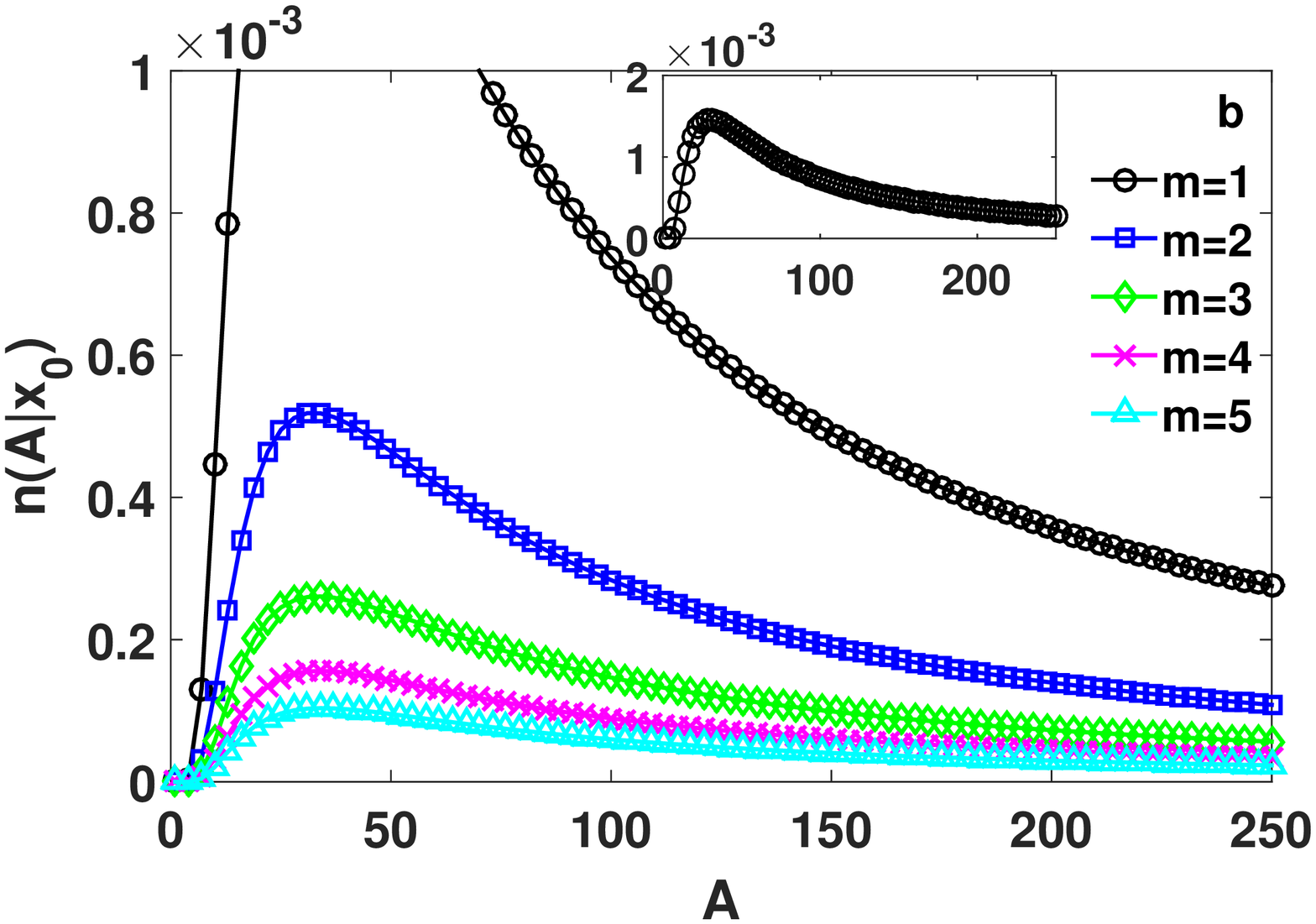}
 \includegraphics[width=2.5in,height=1.25in]{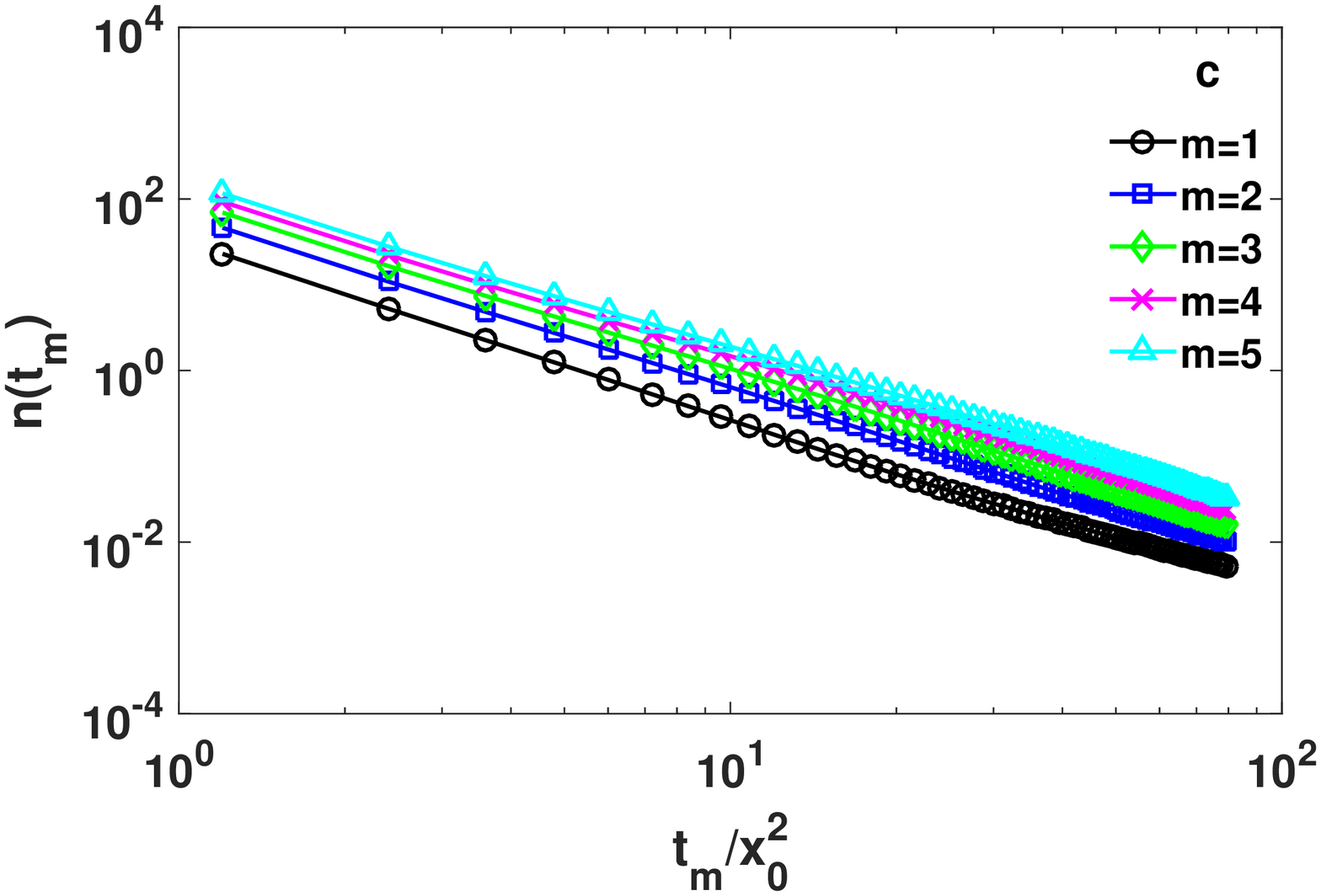}
 \includegraphics[width=2.5in,height=1.25in]{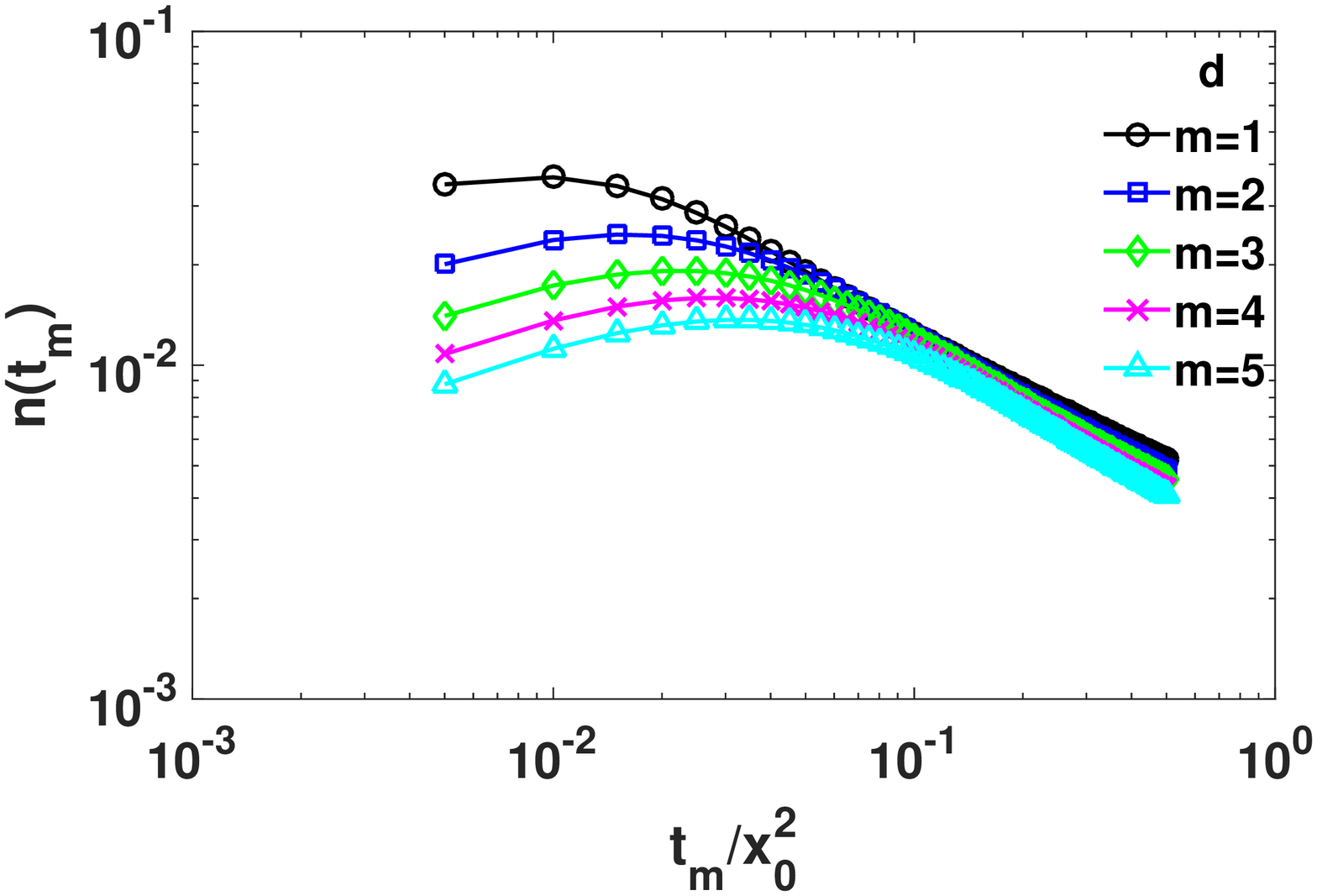}
 \caption{(color online) (a) Plot of first passage time distribution of density profile of excited state $n(t_f|x_0)$  for different mass values which are indicated in the legends and  (b) the distribution of area under the curve, $n(A|x_0)$ which give us information about reactivity of process  is plotted for different mass values.(c) Plot of pdf of $n(t_m)$ before first passage time $t_f$ for the large $t_m$ cases in the log scale. The same thing is plotted for small $t_m$ case in (d) }
\end{center}
\end{figure*}
}
\begin{document}
%opening
\title{Barrierless reaction kinetics :Inertial effect on different distribution functions of relevant Brownian functionals}
\author{ Ashutosh Dubey$^1$, Malay Bandyopadhyay$^1$,and A. M. Jayannavar$^{2,3}$}
\affiliation{1. School of Basic Sciences, Indian Institue of Technology Bhubaneswar, Bhubaneswar, India 751007\\
2. Institute of Physics, Sachivalaya Marg, Sainik School PO, Bhubaneswar, India, 751005.\\
3.Homi Bhabha National Institute, Training School Complex, Anushakti Nagar,
Mumbai-400085, India}
\begin{abstract}
	We investigate effect of inertia on barrierless electronic reactions in solution by suggesting and examining different probability distribution functions (PDF) of relevant Brownian functionals  associated with the lifetime and reactivity of the process. Activationless electronic reaction in solution can be modelled as a free Brownian motion with inertial term in the underdamped regime. In this context we suggest several important distribution functions that can characterize the reaction kinetics. Most of the studies on Brownian functional which has  vast potential application in diverse fields, are confined in the overdamped regime. To the best of our knowledge, we are attempting first time to incorporate the much important inertial effects on the study of different PDFs related with Brownian functionals of an underdamped Brownian motion with time dependent drift and diffusion coefficients using celebrated backward Fokker-Planck and path decomposition methods.  We have explored nontrivial scaling behaviour of different PDFs and calculated explicitly the critical exponents related with the asymptotic limits in time.
\end{abstract}
\pacs{05.40-a, 05.20.-y, 75.10.Hk}
\maketitle
 The picosecond and subpicosecond laser spectroscopy experiments open the doorway to investigate several chemically and biologically important reactions in solution in which the reactants do not face any barrier to the motion along the reaction coordinate \cite{a,b,c}. The dynamics of these reactions demonstrate various interesting behavior such as  the dependence of relaxation rate on the solvent viscosity, the solvent polarity, the temperature, and the wavelength of the exciting light \cite{a,b,c}. The dynamics of these barrierless reactions  obviously differ significantly from traditional chemical reactions where reactants face activation barrier \cite{a,b,c}. In general barrierless reactions are very fast and recent advance spectroscopic techniques enable us to observe the complex motion of the reacting molecules from the reactants to transition state to products over timescales ranging from 10 fs to 10 ps \cite{a,b,c}. An important example of such barrierless reaction is the vision transduction process, which involves a barrierless cis to trans transition of rhodopsin \cite{d}. Other important examples of barrierless reactions are isomerization of stillbene and diphenyl butadiene in solution \cite{c} and nonradiative decay of triphenyl rings \cite{oster} and many more \cite{a,b,c}. The fundamental time scales of such reactions are dictated by the nuclear or electronic rearrangements in the transition state. These barrierless reactions are simplest to model theoretically and open
the doorway to observe directly the motion of a reacting
molecule along the reaction coordinate. In such barrierless
reactions, the solvent friction is the only restriction
to the reaction. The first theoretical treatment on barrierless
reactions was proposed by Oster and Nishijima \cite{oster}
. Later on a systematic studies on such reactions were
made by Bagchi et. al. \cite{bagchi1,bagchi2}. But, all these studies
are restricted in the overdamped regime. One of us \cite{amj1,amj2,amj3} extended this model to the low viscosity or inertial regime. In fact, for small molecules and over short times
(typically less than the friction coefficient
), the inertial motion influence on reaction rates is significant. It
is shown that the transient behaviour of the integrated
excited state population is very much sensitive to initial
velocity profile of the reaction coordinate and may
cause to a nonlinear dependence of viscosity on the reaction
rate \cite{amj1,amj2}. In this model the motion of a reactant
molecule on the reaction coordinate is modelled by a dynamical
Langevin equation \cite{e,f}
\begin{equation}
\ddot{x}=-\gamma\dot{x}+\eta(t),
\end{equation}
where, x is the reaction coordinate of the molecule, m
is the mass of the molecule, $\gamma$ is the friction coefficient
of solvent and the associated white Gaussian noise is denoted
by $\eta(t)$. The gaussian noise $\eta(t)$ and the associated friction coefficient $\gamma$ will maintain the following fluctuation-dissipation relation
\begin{eqnarray}
<\eta(t)>&=&0 \nonumber \\
<\eta(t)\eta(t^{\prime}>&=&\frac{2K_BT\gamma}{m}\delta(t-t^{\prime})
\end{eqnarray}
\indent
where, $< - >$ denotes the ensemble average over all the
possible realizations of the random force $\eta(t)$, $T$ is the temperature of the heat bath
and $k_B$ is the Boltzmann constant.\\
 %Recent analysis by Bodrova et. al \cite{bodrova1,bodrova2}, shows that for certain underdamped scaled Brownian motion (UDSBM) the overdamped limit fails to describe the long time behaviour of the system and may not exist at all for certain parameter regimes. The continual inertial effects play a significant role even at significantly long times. So, question may arise regarding the applicability of the overdamped limit to describe the long time behaviour of such anomolously diffusing particle which are observed in living biological cells and complex fluids \cite{metzler1,barkai,meroz}, submicron lipid \cite{jeon}, insulin granules \cite{tabei} and many more systems \cite{pan,weiss}. In the present paper, we consider such kind of Brownian motion as mentioned in Eq. (1) with time dependent drift and diffusion coefficients in the context of barrierless electronic reaction kinetics \cite{amj1,amj2}. Most importantly, we show continual inertial effect on several pdfs which characterize the barrierless reactions and they are derived analytically with Brownian functional mecahnism of a Brownian motion which can describe such barrierless electronic reaction kinetics.\\
\indent
Consider a Brownian motion which can be represented by a typical path $x(\tau)$. Thus, a Brownian functional is defined in the interval $\lbrack 0,t \rbrack$ as $T=\int_{0}^{t}U(x(\tau))$. Here, $U(x(\tau))$ is some prescribed arbitrary function. For each realization of Brownian path, the quantity $T$ will be different and it is of great interest to study the pdf of $T$.In this respect several interesting questions of wide inter
est can be raised such as, (i)the probability of finding the
system in a certain domain at a certain instant (survival
probability), (ii)the pdf of time $P(t_f|x_0)$ at which the
system exit a certain domain first time (known as first
passage time $t_f$ ) starting from initial point $x_0$, (iii)the
pdf $P(M)$ of the maximum value of a BM process before of its first passage time, and (iv)the joint probability distribution $P(M; tm)$ of the maximum value M and its occurrence time $t_m$ before the first passage time of the BM process. Brownian functionals are extensively used in the study of diverse fields ranging from probability theory \cite{snm1,snm2} and finance \cite{snm1} to disordered systems and mesoscopic physics \cite{snm1}.  \\
\indent
All the above mentioned PDFs are calculated and discussed for simple Wiener and Ornstein-Uhlenbeck processes \cite{snm1,snm2,snm3} as well as in the context of DNA breathing dynamics \cite{malay}. But, all these discussions are based on constant drift and diffusion terms. More importantly, all these discussions are restricted in the overdamped or high friction limit. However, the extension to a time dependent drift and time dependent diffusion terms for an underdamped Brownian motion is hard to solve. This is mainly because of the fact that the system has broken both the space and time homogeneity.\\
\figOne
\indent
The main objective of the present work is to demonstrate
inertial effect in different PDFs which can characterize such barrierless reactions using well studied \cite{snm1,snm2,snm3} Brownian functionals method. To
the best of our knowledge, it is the first attempt to incorporate
inertial effect in first passage study which is one of
the important unsolved problem. The other objective of
this work is the extension of the use of the recently studied
backward Fokker-Planck (BFP) method \cite{snm1} and the
path decomposition (PD) method \cite{snm1} in the underdamped regime or inertial regime of a Brownian motion with time dependent drift and diffusion coefficients. Both the BFP
and PD methods are based on the Feynman-Kac formalism
\cite{kac} and both of them are first time used for exploring
underdamped BM process with purely time dependent
drift and diffusion terms for barrierless intramolecular
electron-transfer reactions. Both the techniques are
extensively used in studying many aspects of classical
Brownian motion, as well as for exploring different problems
in computer science and astronomy \cite{snm1,snm2,snm3}. For
the first time, we consider these elegant methods to study
the Brownian functionals for a BM with purely time dependent
drift and diffusion in the context of barrierless
reaction kinetics.\\
\indent
{\bf Model,Methods and Measures:} One of the most direct
way to demonstrate the effect of the solvent friction
on a reaction in solution is to study a reaction without
an activation barrier. Several studies have been made
on barrierless intramolecular electron transfer reaction
using various spectroscopic techniques \cite{b}.  Equation (1)
can be cast into the Kramers equation for the full phase
space probability distribution function $P(x, v, t)$, where
v is the velocity of the particle. The full phase space
Kramers equation is given by
\begin{equation}
\frac{\partial P}{\partial t}=\frac{\gamma}{m}\frac{\partial}{\partial v}\Big(v+\frac{k_BT}{m}\frac{\partial}{\partial v}\Big)P-v\frac{\partial P}{\partial x},
\end{equation}
The fundamental solution of the conditional probability
$P(x,v,t |x_0, v_0, t = 0)$ of the Kramers equation is well
known and its expression is too long to discuss here \cite{e}.
Now, considering this solution one can easily obtain an
expression for the number density $n(x, t)$ by integrating
out the velocity component \cite{amj2}
\begin{equation}
n(x,t)=\frac{1}{\sqrt{2\pi F(t)}}\exp\Big(-\frac{\lbrack x-x_0-\frac{v_0}{\beta} (1-e^{-\beta t})\rbrack^2}{2F(t)}\Big),
\end{equation}
where $F(t)=q\beta^{-3}(2\beta t-3+4e^{-\beta t}-e^{-2\beta t}$, $\beta=\frac{\gamma}{m}$, $q=\frac{\beta k_BT}{m}$. The associated Fokker-Planck equation
for n(x; t) can be obtained as follows \cite{amj2}
\begin{equation}
\frac{\partial n(x,t)}{\partial t}=D(t)\frac{\partial^2 n(x,t)}{\partial x^2}-\mu(t)\frac{\partial n(x,t)}{\partial x},
\end{equation}
The projected evolution of $n(x,t)$ is non-Markovian in nature,
since the evolution of $n(x,t)$ is dependent on the initial
velocity $v_0$ which arises from the non-Markovian process.
It can be easily observed that Eq. (5) easily reduces to
Smoluchowski equation for a free particle in the long time
limit $\beta t >> 1$. In this context, the four PDFs $P(t_f|x_0)$, $P(A|x_0)$, $P(M)$ and $P(M,t_m)$ are of great interest for determining different properties of the excited state. The first passage time pdf $P(t_f|x_0)$ will provide us about the lifetime of the excited state. A related quantity is the survival probability $C(x_0,t)=1-\int_{0}^{t}P(t_f|x_0)dt_f$ which can be inferred from the experiments by using different spectroscopic technique. For the path as described by Eq. (1), one can introduce the area under the path before first passage time as $A=\int_{0}^{t_f}x(t^{\prime})dt^{\prime}$ and calculate its pdf $P(A|x_0)$. This quantity provide us the information about the effective reactivity of the barrierless reaction process. Another proposed measureable quantity for quantifying the reactivity of barrierless reaction is the distribution of the maximum number of reactants in the excited state before its first passage time, i.e., $P(M)$. Since, the timescale of this reaction is very short a relevant measure for testing the reactivity is its maximum number of reactants in the excited state before its decay. Finally, the joint probability distribution function $P(M,t_m)$ will provide the information about both the maximum reactivity and its occurrence time.\\
\indent
Transforming from $(x,t)$ to $(z,\tau)$ by using the following transformation equations
\begin{equation}
\tau=\int D(t)dt +A
\end{equation}
and
\begin{equation}
z=x+\int\mu(t)dt +B
\end{equation}
one can convert Eq. () in $(z,\tau)$ space as follows :
\begin{equation}
\dfrac{\partial n(z,\tau)}{\partial \tau}=\dfrac{\partial^{2}n(z,\tau)}{\partial z^{2}}
\end{equation}
PDF of Brownian functional over a fixed time interval $[0,\tau_{f}]$ is defined as
\begin{equation}
T=\int_{0}^{\tau_{f}}U(z(\tau))d\tau,
\end{equation}
and our aim is to calculate PDF $P(T|z_{0})$. Using backward Fokker-Planck method as discussed in details in Refs. \cite{snm1,snm2} one can obtain :
\begin{equation}
\dfrac{d^{2}Q}{dz_{0}^{2}} -pU(z_{0})Q(z_{0})=0
\end{equation}
Substituting $U(z_{0})= 1$ and using proper boundary conditions $Q(z_0=0)=1$ and $Q(z_0\rightarrow\infty)=0$ we obtain :
\begin{equation}
n(\tau_{f}|z_{0})=\dfrac{z_{0}}{\sqrt{4\pi}}\dfrac{e^{-z_{0}^{2}/4\tau_{f}}}{\tau_{f}^{3/2}}
\end{equation}
%\figTwo
Then going back to the original variables $(x,t)$ we can obtain the first passage time distribution of the number density in the excited state :
\begin{eqnarray}
&&n(t_{f}|x_{0},v_0)=\beta \sqrt{\dfrac{2\beta}{q\pi}}\dfrac{(e^{-\beta t_{f}}-1)^2[\beta x_{0}+v_{0}(1-e^{-\beta t_{f}})]}{(2\beta t_{f}+4e^{-\beta t_{f}}-e^{-2\beta t_{f}}-3)^{3/2}}\nonumber \\
 &&\exp\bigg(-\dfrac{\beta[\beta x_{0}+v_{0}(1-e^{-\beta t_{f}})]^2}{2q(2\beta t_{f}+4e^{-\beta t_{f}}-e^{-2\beta t_{f}}-3)}\bigg)
\end{eqnarray}
We can also extract much more information. Another important quantity is the survival probability of the population remaining at the excited state surface at time t after excitation :
\begin{eqnarray}
  C(x_0,v_0,t) &=&1-\int_{0}^{t} dt_f n(t_f|x_0,v_0) \nonumber \\
   &=& 1-\frac{1}{\Gamma(1/2)}\Gamma(3/2,\frac{z_0^2}{4\tau},
\end{eqnarray}
with $z_0=x_0+\frac{v_0}{\beta}(1-e^{-\beta t})$ and $\tau =\frac{q}{2\beta^3}(2\beta t +4e^{-\beta t}-e^{-2\beta t}-3)$. This quantity can be tagged with the experiments. The distribution of $A$ can be computed from Eq. () by substituting $U(z_{0})=z_{0}$. Using proper boundary conditions as mentioned above one can show:
\begin{equation}
Q(z_{0})=3^{2/3}\Gamma(2/3)Ai(2^{1/3}p^{1/3}z_{0})
\end{equation}
Taking the inverse Laplace transform
\begin{equation} P(A(\tau_{f})|z_{0})=\dfrac{2^{1/3}}{3^{2/3}\Gamma(1/3)}\dfrac{z_{0}}{[A(\tau_{f})]^{4/3}}exp\bigg[-\dfrac{2z_{0}^3}{9A(\tau_{f})}\bigg]
\end{equation}
Now, going back to original variables one can obtain the distribution function of A :
\begin{eqnarray}
&&n(A|x_{0},v_0)=\dfrac{2^{1/3}}{3^{2/3}\Gamma(1/3)}\dfrac{q\beta^{-3}(e^{-\beta t_{f}}-1)^2}{[A]^{4/3}} \nonumber \\
 &&\times[\beta x_{0}+v_{0}(1-e^{-\beta t_{f}})]\exp \bigg[-\dfrac{2[\beta x_{0}+v_{0}(1-e^{-\beta t_{f}})]^3}{9\beta^3 A}\bigg]\nonumber \\
\end{eqnarray}
Using well known path decomposition method \cite{snm1,snm2,snm3}, one can find the joint probability density of density in the excited state,$n(M,t_m)$ which can provide both the information : (i)the maximum number of particles available in the excited state before its decay ($t_f$) as well as the (ii)the time at which it will attain its maximum before first passage time or decay :
\begin{eqnarray}
&&n(M,t_{m})=\dfrac{q\pi (e^{-\beta t_{m}}-1)^2}{\beta^2 M^3}\sum_{p=1}^{\infty}(-1)^{p+1}p\nonumber \\ &&\times\sin\bigg[\dfrac{p\pi[\beta x_{0}+v_{0}(1-e^{-\beta t_{m}})]}{\beta M}\bigg]\nonumber \\
&&\times\exp\bigg[\dfrac{p^2\pi^2 q(2\beta t_{m}+4e^{-\beta t_{m}}-e^{-2\beta t_{m}}-3)}{4M^2\beta^3}\bigg]
\end{eqnarray}
The closed form expressions of  marginal distribution $n(t_m)$ can also be obtained from Eq. (16) by integrating out $M$ in the following two limits (i)$t_m/x_0^2 >> 1$ and (ii)$t_m/x_0^2 << 1$. In the large $t_m$ limit one can find :
\begin{widetext}
\begin{equation}
n(t_{m}) \approx\log2~ \beta\sqrt{\dfrac{\beta}{q\pi}}\dfrac{(e^{-\beta t_{m}}-1)^2~[\beta x_{0}+v_{0}(1-e^{-\beta t_{m}})]}{[(2\beta t_{m}+4e^{-\beta t_{m}}-e^{-2\beta t_{m}}-3)]^{3/2}}
\end{equation}
\end{widetext}
and the small $t_m$ limits give us
\begin{widetext}
\begin{equation}
n(t_{m})=\sqrt{\dfrac{q\beta}{\pi}}\dfrac{(e^{-\beta t_{m}}-1)^2}{[\beta x_{0}+v_{0}(1-e^{-\beta t_{m}})][(2\beta t_{m}+4e^{-\beta t_{m}}-e^{-2\beta t_{m}}-3)]^{1/2}}.
\end{equation}
\end{widetext}
All the four $PDF$s are plotted in figure 1. The basic message as obtained form the figures is that the mass effect is distinctly observed in the transient regimes. As we go to the asymptotic limit in time we can observe universal scaling behaviour of all four PDFs. We observe nonmonotonic behaviour of the pdfs of first passage time $P(t_f|x_0)$ and area $P(A|x_0)$. In the low-friction regime, both $P(t_f|x_0)$ and $P(A|x_0)$ increase with time to reach a maximum and then crosses over to universal scaling behaviour regime as observed by an overdamped Brownian particle. In this time asymptotic limit, we observe that (i)PDF of first passage time $P(t_f|x_0)\sim t_f^{-3/2}$, (ii) $P(A|x_0)\sim A^{-4/3}$,  (iii)$P(M)\sim M^{-2}$ and (iv)$P(t_m)\sim t_m^{-3/2}$. But, the transient regime shows nonuniversal behaviour (depends on mass) which is mainly due to the inertial term of Eq. (1). $P(t_m)$ has power law behaviour at small tails also but the behaviour is nonuniversal and it depends explicitly on mass. We also show that in the low friction regime all four PDFs depends crucially on the initial conditions,i.e., on the velocity ($v_0$) and position ($x_0$). In the present context, we consider special condition of $p(v_0)=\delta(v_0)$ and $p(x_0)=\delta(x-1/2)$, which corresponds to a particle placed at the mid point with zero velocity initially. One may consider different initial conditions and different excited state potential for futher research. \\
It is quite generally assumed that the overdamped Langevin model provides a very good description of the barrierless reaction kinetics \cite{bagchi1,bagchi2,oster}. We establish and investigate an anomalous diffusion process which governed by an underdamped Brownian motion with an explicit time dependence of the diffusion and drift coefficients to capture correct characteristics of barrierless reaction kinetics. The four PDFs distinctly show the persistent inertial effects and it plays a non-negligible role in the transient regime as well as at reasonably long time. This multifaceted problem help us to investigate inertial effect to characterize barrierless reactions in the low friction regime. On the other hand, the present study helps to make advancement in the Brownian functional method to a problem of Brownian motion with an explicit time dependence of the diffusion and drift coefficients in thelow friction or inertial regime. Our investigation will be helpful in understanding fundamental time scales related with barrierless reactions which are dictated by the nuclear or electronic rearrangements in the transition state.
\begin{acknowledgments}
MB acknowledge the financial support of IIT Bhubaneswar through seed money project SP0045. AMJ thanks DST, India  for award of J C Bose national
fellowship.
\end{acknowledgments}

\end{document}